\documentclass[sigconf]{acmart}

\copyrightyear{2024}
\acmYear{2024}
\setcopyright{acmlicensed}\acmConference[FDG 2024]{Proceedings of the 19th International Conference on the Foundations of Digital Games}{May 21--24, 2024}{Worcester, MA, USA}
\acmBooktitle{Proceedings of the 19th International Conference on the Foundations of Digital Games (FDG 2024), May 21--24, 2024, Worcester, MA, USA}
\acmDOI{10.1145/3649921.3659843}
\acmISBN{979-8-4007-0955-5/24/05}

\AtBeginDocument{%
  \providecommand\BibTeX{{%
    \normalfont B\kern-0.5em{\scshape i\kern-0.25em b}\kern-0.8em\TeX}}}

\copyrightyear{2024}
\acmYear{2024}
\setcopyright{acmlicensed}\acmConference[FDG 2024]{Proceedings of the 19th International Conference on the Foundations of Digital Games}{May 21--24, 2024}{Worcester, MA, USA}
\acmBooktitle{Proceedings of the 19th International Conference on the Foundations of Digital Games (FDG 2024), May 21--24, 2024, Worcester, MA, USA}
\acmDOI{10.1145/3649921.3659843}
\acmISBN{979-8-4007-0955-5/24/05}

\usepackage{xspace}
\usepackage{caption}
\usepackage{subcaption}

\usepackage[clock]{ifsym}

\acmConference[FDG '24]{}{May 21--24,
  2024}{Worcester, USA}
\newcommand*{\mathabxbfamily}{\fontencoding{U}\fontfamily{mathb}\selectfont}
\DeclareFontFamily{U}{mathb}{\hyphenchar\font45}
\DeclareFontShape{U}{mathb}{m}{n}{
      <5> <6> <7> <8> <9> <10> gen * mathb
      <10.95> mathb10 <12> <14.4> <17.28> <20.74> <24.88> mathb12
      }{}

\newcommand*{\Neptune}{{\text{\mathabxbfamily\char"48}}}

\makeatletter
\def\@fnsymbol#1{\ensuremath{
\ifcase#1\or
\Small{\text{\VarTaschenuhr}} \or  
\Neptune \or 
\mathsection \or
\mathparagraph \or
\|\or **
\or
\dagger\dagger
\or \ddagger\ddagger
\else\@ctrerr\fi}}
\makeatother

\renewenvironment{quote}{%
  \list{}{%
    \leftmargin0.5cm   
    \rightmargin\leftmargin
  }
  \item\relax
}
{\endlist}

\begin{document}

\title{Permalife Of The Archive: Archaeogaming As Queergaming
}


\author{Florence Smith Nicholls}
\affiliation{%
  \institution{Queen Mary University of London}
  \city{London}
  \country{UK}
}
\email{florence@knivesandpaintbrushes.org }

\authornote{This author should be cited with their full surname "Smith Nicholls" and they/them pronouns.}

\renewcommand{\shortauthors}{Smith Nicholls}
\newcommand{\nbr}{\textit{Nothing Beside Remains}\xspace}

\begin{abstract}
Archaeogaming and queer games studies have both grown as para\-digms in the last decade. The former broadly refers to the archaeological study of games, while the latter concerns the application of queer theory to the medium. To date, there has been limited engagement of archaeogamers with queer games scholarship, and vice versa. This article argues that there are epistemological parallels between the two; they are both concerned with the limits and ethics of representation, the personal and political contexts of game development and engagement with video games through transgressive play. 

The paper is structured around an extended literature review and three vignettes that reflect on the author’s personal experience of conducting archaeogaming research; 
an ethnographic study of \textit{Wurm Online}, an archaeological survey of \textit{Elden Ring} and a player study of the generative archaeology game \textit{Nothing Beside Remains}. While archaeogaming can learn from the centring of subjective lived experience and labour in the queer games sphere, archaeogaming as a form of game preservation can also benefit queer games studies. 

\end{abstract}





\maketitle

\section{Introduction}

As the portmanteau would suggest, archaeogaming is the archaeological study of video games. The term was first coined in 2013 by Andrew Reinhard \cite{reinhard2013}, who was influenced by Emily Johnson’s work \cite{johnson2013} \cite{johnson2013experienced} on ethnographies of player engagement with the past in \textit{Skyrim}. Earlier work had also explored the representation of heritage in video games, however Johnson’s work is notable not just for considering the archaeological implications of a game with a fictional fantasy setting, but also engaging with the nascent flourishing of archaeogaming discussions on social media. 
This mantel was taken up a decade later with Politopoulos et al’s survey of playful approaches to archaeology, in which they identify grassroots blogging in the 2010s as being a key conduit through which archaeogaming developed \cite{politopoulos2023finding}. One example is Dennis’ blog \textit{Gingerygamer} (sadly no longer accessible), which provided a particularly succinct definition of the emerging field:
\begin{quote}... the utilization and treatment of immaterial space to study created culture, specifically through video games \cite{dennis}
\end{quote}
As will be elaborated on below, very broadly, archaeogaming can be defined as the study of how heritage is represented in games, the creation of archaeological games, and the application of archaeological theory and method in recording games as archaeological sites. Histories of the field can also be found in \cite{rassalle2021archaeogaming} and more recently \cite{reinhard2024practical}.

Queergaming, and queer games studies, are also resistant to discrete categorisation. Again, broadly speaking, queer games studies can be understood as the intersection of queerness and games. The term "queergaming" was coined by Edmond Y. Chang \cite{chang2017queergaming}, a provocation for queer games beyond representation but also in terms of design, play and remediation. Like archaeogaming, it has antecedents in earlier scholarship \cite{cassell2000barbie}, but really started to emerge as a paradigm in the last decade. Ruberg \cite{rubergfun}, one of the most prolific queergaming scholars, points out that, like Chang, that the paradigm draws from two major definitions of ‘queer’: as an umbrella term for LGBTQA+ identities, and as a concept referring to a different way of being in the world. Also like archaeogaming, queergaming, then, refers to various different ways to approach games beyond representation:
\begin{quote}
It refigures games as systems of pleasure, power, and possibility, excavating the queer potential that can be found in all games \cite{rubergqueer2017}
\end{quote}
There has been limited archaeogaming scholarship that draws from queer theory. In 2018 I presented a paper on "Archaeogaming as Queergaming" at the Computer Applications and Quantitative Methods in Archaeology conference \cite{florencequeergaming}, which laid the ground work for this paper. There are a handful of pieces that discuss the representation of queer characters and archaeology in games \cite{meier2022hardest} \cite{nicholls2022androgynous}. There is more work that considers the queerness of digital archaeology more broadly. Morgan \cite{morgan} has explored the potential of a cyborg archaeology that engages with the complications of digital practice and embodiment, with reference to the subversive potential of queerness beyond singular narratives. Cook \cite{cookembodiying2019} advocates for a digital archaeology that destabilises traditional hierarchies of knowledge, citing the archaeology dating game \textit{C 14 Dating} as a rare example of queer archaeologist representation. This article aims to build on this foundational work in drawing parallels between archaeogaming and queergaming in the hope of demonstrating that not only have games always been queer, archaeologies of games have always been queer. 

The remainder of the paper is organised as follows: in \textit{Background and Related Work} I provide an extended literature review of relevant archaeogaming and queergaming scholarship; \textit{Motivation} provides further context for the work; in \textit{Vignettes} I reflect on my personal experience of conducting three archaeogaming projects; and in \textit{Discussion} I tease out the themes of these vignettes and how they relate to queergaming.

\section{Background and Related Work}

\subsection{Representation and ethics}
\begin{quote}
...adventure games are tainted by the ‘Indiana Jones’ quandary. Archaeology is glorified via popular culture, but not for preservation, only for exploration of novelty and the demonisation and destruction of other cultural perspectives. \cite{champion2004indiana}
\end{quote}

\noindent This quote from a 2004 article by Champion succinctly encapsulates some of the main concerns that archaeogaming grapples with in terms of how archaeology, heritage and archaeologists themselves are represented in video games. Hageneuer \cite{hageneuerarchaeogaming2021} identifies three main themes that continue to be perpetuated through problematic depictions of archaeologists: imperialism, racism and sexism. Lara Croft, probably the most famous video game archaeologist, is a symbol of female objectification,  while also perpetuating the white saviour trope. Concerns over ethical representations of archaeology also extend to looting mechanics that proceduralise an extractivist and imperialist approach to the past \cite{dennis2016archaeogaming}. Meghan Dennis has been one of the strongest proponents of taking the ethical implications of archaeological misrepresentations in games seriously. In her doctoral thesis she presents the results of a survey into player perceptions of archaeologists based on their depictions in games, finding that players will be more impressionable if they have no experience of archaeology outside of that context \cite{dennis2019archaeological}.

Other archaeogaming scholarship relating to representation focuses on period-specific settings in the medium. The \textit{Assassin’s Creed} franchise is particularly well-represented in this work \cite{caseyassassins2021, poironassassins2021}. Bowman et al explore how perceived realism affects enjoyment of titles in the franchise \cite{bowmananimating2023}. Other work also reflects on the pedagogical potential of archaeogaming, and particularly the \textit{Assassin's Creed} games, in the classroom \cite{martino2023archeogaming}. Politopoulos et al \cite{politopouloshistory2019} make the point that its important to engage with these AAA titles in particular, as they are likely to be a common entry point to the study of the past for many students. Hanussek \cite{hanussek2019conducting} raises concerns over Ubisoft holding private ownership of a Notre Dame cathedral model; studios and franchises having a cultural monopoly in this space is a topic that deserves much further consideration. 

Mol et al conducted a survey \cite{politopoulos2021video} of games on Steam with the “historical” tag, finding that strategy games were over-represented. This quantification of games containing historical representation is not dissimilar to the quantification of games with queer representation by Utsch et al \cite{utschqueer2017}. Among their other findings, they found that LGBTQA+ representation is concentrated in RPGs, adventure and action games. Shaw et al have also conducted a survey of more than 300 games spanning 30 years, stressing the importance of looking at queer diversity in relationships, gender, and even location and artefacts, as well as sexuality \cite{shawwhere2016}. They also make the point that, as upsetting as it is to encounter, it is important to document incidents of homophobia and transphobia in games: 
\begin{quote}
Casual homophobia and transphobia in game texts demonstrate that it is not difficult for game makers to see the relevance of LGBTQ content in these texts. The challenge moving forward is to see if casual inclusivity is as possible as casual offensiveness. \cite{shawwhere2016}
\end{quote}
There is considerable scholarship on case studies of queer representation in games; one of particular relevance to the above point is Adams’ \cite{adams2018bye} work on heroic androgyny and gender variance in games: masculine-leaning gender androgyny is celebrated as aesthetically pleasing in the case of Link’s character in \textit{The Legend of Zelda} franchise, while a trans woman was used as a punchline in \textit{Breath of the Wild}. Cole et al reference a survey, via the charity \textit{Represent Me}, of 1,237 participants in which 150 stated that “realism” was important to them in terms of queer representation \cite{colerepresentationsnodate}. Even when queerness is represented or queer relationships are offered as an option, presenting a uniformly tolerant world has been criticised on the basis that it reflects a “liberal logic of sameness.” \cite{greerplaying2013} Ruberg also identifies the limits of pushing for more queer representation when there are cases of the wider gaming community straight-washing queer characters in \textit{Undertale}, for example \cite{rubergundertale}.

While comparisons can be drawn between the concerns and limits of representations with regards to archaeology and queerness in video games, it would clearly not be fair to equate these too. Being an archaeologist is a professional identity, being queer is a marginalised one.

\subsection{Game development}

Within the existing archaeogaming literature, there is considerable discussion of developing games as pedagogical tools, and game development itself as a learning exercise. Champion \cite{champion2020games} reflects on his PhD project in which he evaluated 80 students playing a browser-based recreation of the Mayan city Palenque. He reports that those participants who completed tasks more quickly in the game also scored lower in terms of memory recall when asked about Mayan history, which is a good indication of the brittleness of these kinds of applied game metrics. Hiriart \cite{hiriart2020good} asked primary school children to draw how they imagined Anglo-Saxon life before and after playing a game set in the period, reflecting that a deeper understanding of history can only be accessed through empathising with people in the past. McKinney et al created a multi-component digital kit for young people learning in both formal and informal environments, with an emphasis on the importance of fostering emotional experiences, in that:
\begin{quote}
“The evidence suggests that it is through personal, emotional connections that humans are more likely to be primed to acquire knowledge about, protect and promote the archaeological record” \cite{mckinney2020developing}.
\end{quote}
Archaeogaming can be understood as a form of public archaeology, or public engagement \cite{ezzeldin}. This can be in the form of game development, and teaching others how to make games as a pedagogical process in of itself. The affordances of Twine as an open source game-making tool for archaeologists that favours non-linear storytelling has been discussed by Tara Copplestone \cite{copplestone2017}, and as a challenge for scholars and students to engage with their work in a form of non-traditional dissemination \cite{value}. Archaeogamers have also reflected on their own creative and interpretive process in making games and digital reconstructions. Kingsland \cite{kingslandarchaeogaming2023} explicitly refers to her own work as generating serious games, emphasising how narrative and environment design were core to her work on re-using digital archaeological datasets of the Roman Villa del Casale. Morgan \cite{morgan2009re} contends that the process of reconstructing the site of Çatalhöyük in \textit{Second Life} “has increased my engagement with the materiality of the objects and how they might have related to each other during their use-lives.”

In some cases, archaeologists have used existing tools or mods as part of their development process. Majewski \cite{majewski2017} contends that there are four types of cultural heritage games; commercial games, serious games, culture-centric games and mods. Regardless of the limitations of this model, Majewski’s point about mods being an accessible entry point to game development to a wider audience of creatives and archaeologists is important.  This also chimes with Graham’s \cite{graham2020} provocations on mods as a form of resistance against hegemonic structures, as we must consider the labour ethics of game development. This tracks with Ashlee Bird’s ROM hack of \textit{Super Mario Bros}. which replaced Mario with the figure of Gluskabe, an Abenaki hero, as: 
\begin{quote}
“Indigenous authors of digital games push back against the narrative of the people trapped in time.” \cite{bird2021synthetic} \end{quote}
Archaeological games can be about the recent past, and also personal histories, as was the case with the short narrative game \textit{Fragments of Him}. Mata Haggis-Burridge describes the game, which was discussed at the Interactive Pasts conference \cite{haggis2021personal}, as a recent-history period drama based on many real life conversations that he had when coming out as bisexual in the 1990s. This focus on personal lived experience and creative projects is also a key concern in writing on queer game development. As Anna Anthropy puts it in \textit{Rise of the Videogame Zinesters} when she dreams of a future in which game development is accessible to a wider remit of people:
\begin{quote}
“...even if a game isn’t original, it’s personal, in the way a game designed to appeal to target demographics can’t be. And that’s a cultural artefact our world is a little richer for having” \cite{anthropy}
\end{quote}
Anthropy is passionate about tools, like the aforementioned open source Twine, that have been taken up by developers from non-traditional coding backgrounds, especially queer people. Furthermore, in discussing the so-called ‘Twine revolution,’ Harvey \cite{harvey2014twine} references "personal games" of the early 2010s which were often made by one person and depicted autobiographical experience. However, Harvey also cautions that the innovative work produced by marginalised groups may be co-opted by the mainstream games industry, depoliticising it outside of the original context of production. Furthermore, Freedman \cite{freedmanengineering2018} warns against a binary opposition between tools like Twine and proprietary engines, arguably advocating for a media archaeology approach that traces the potential representational politics of code beyond simple output, considering more fundamental logics that underpin software architecture. More broadly, Aceae and Brewster have recognised the potential of queer games to destabilise the liberal, capitalist logics of both academia and the games industry \cite{aceaerising2022}.
Bo Ruberg’s \textit{The Queer Games Avante-Garde} \cite{rubergavante2020} is a collection of interviews with queer game developers from different backgrounds, thus an essential resource for understanding different perspectives on queer game making. Several interviews counter the focus on concepts of gamification and empathy in the archaeogaming literature. For example, Andi McClure \cite{mcclure} believes that “gamification misses the point. Games are a large number of things, not just their incentive structures.” Liz Ryerson also rejects the pressure to perform her identity through the games she makes, and laments that:
\begin{quote}
“It’s those “empathy games” that get noticed. My work isn’t like that so it doesn’t get noticed in the same way” \cite{ryerson}
\end{quote}
In looking at how the archaeogaming and queergaming communities have been represented in the academic literature, it is interesting to note that although archaeogamers write from a position of professional identity, they do not tend to engage with the labour politics of the games industry to the extent that the queer games community does. Perhaps as a result of lived experience and personal subjectivity being so core to the queer game-making space, the conversation seems to have moved beyond empathy, while the archaeogaming literature values empathy with past peoples.

\subsection{Methodologies and mechanics}

A third core strand under the archaeogaming umbrella is the application of archaeological methodologies to games in order to better understand and record human (and non-human) activity within them. Andrew Reinhard has been a strong advocate for this approach, in his thesis \cite{reinhard2019archaeology} arguing
\begin{quote}
At the core of my research is the fundamental argument that human-occupied digital spaces can be studied archaeologically with existing and modified theory, tools, and methods to reveal that human occupation and use of synthetic worlds is similar to how people behave in the natural world.
\end{quote}
There have been limited examples of archaeological surveys in games, with Reinhard’s work on \textit{No Man’s Sky} \cite{reinhardarcheology2021} perhaps being the best known. Hansen’s recent work on the abandoned MMO \textit{Star Wars Galaxies}\cite{hansen2022life}, in which he uses a multi-modal approach of archaeological site mapping and anthropological semi-structured interviews, also stands out. Graham’s autoethnography of a \textit{Minecraft} playthrough is also notable for being self-reflexive, in which he poses that in order to be ethical digital archaeologists we must challenge the perceived norms of intended gameplay \cite{graham2020}. There has been some meta-analysis of the specific methodologies that archaeogamers have applied to digital space, such as the use of plans and maps \cite{smithnichollsmeta}. An archaeological survey of player messages and blood stains in \textit{Elden Ring} (discussed as one of the vignettes below) grappled with what constitutes archaeological context in a mingleplayer games in which player created content is largely ephemeral \cite{smith2022dark}. More recently, Reinhard and Zaia have published on using GIS and photogrammetry to record human-occupied game environments, stressing the importance of this methodology given the rapid pace at which these landscapes change \cite{reinhardphotogrammetry2023}. This arguably represents a shift in thinking from Reinhard’s earlier statement that people behave similarly in analogue and digital worlds, also reflecting Morgan’s thoughts that purely skeuomorphic emulation of analogue archaeological methods in a digital context “may inhibit truly transformative uses of these technologies.” \cite{morganskeu}

Related to the topic of the rapid accumulation of human activity in digital games, Aycock calls for a paradigm shift in archaeology to adequately engage with the vast amount of digital artefacts in the contemporary era \cite{aycock2021coming}. 
Building on earlier work into archaeologies of flash drives \cite{moshenska2014} and hard drives \cite{perrymorgan}, Aycock and Biittner have spearheaded work on reverse-engineering games such as Mystery House \cite{aycock2019inspecting}, and worked with Ganesh on a framework that uses AI to reverse engineer historical video games at scale \cite{ganesh}. This work has some echoes in the Digital Ludeme Project, which uses AI to play ancient board games, attempting to fill in the gaps of missing rulesets\cite{crist2023digital}.

The subject of play has not historically been central to archaeology, which Politopoulos et al would like to change \cite{politopoulos2023finding}. Their work on contextualising fun in the archaeological record is laudable, and also presents an interesting foil to queer game studies that make frustrating and uncomfortable play central to their work.  In a 2015 article \cite{ruberg2015no}, Ruberg makes the point that games being “just for fun” was a rallying cry of the GamerGate harassment campaign in 2014, with Zoe Quinn’s \textit{Depression Quest} being a target of vitriol. This point remains relevant in the current games discourse, as a self-styled “GamerGate 2” has been incited over Sweet Baby Inc, a narrative development and consultation studio which specialises in cultural consultation and sensitivity reading \cite{sweetbaby}. In later work Ruberg points out the both “fun” and “empathy” have been employed as buzzwords to legitimise games, however critiquing them both 
\begin{quote}
Is founded on the urge to resist hegemonic beliefs about the value of affect - whether those hegemonies dictate how a player should feel, what social uses a game should have, or who has a right to lay claim to queer feeling \cite{rubergfun}
\end{quote}
The embracing of discomfort and failing in queer games studies can be traced back to the work of queer theorists like Jack Halberstam \cite{halberstam2011queer} and José Esteban Muñoz \cite{munoz2019cruising}, who call for imagining queer futures, even and especially as they do not adhere to heterosexual, capitalist models of success. The embracing of failure is also a key component of Edmund Chang’s definition of queergaming, which includes queer representation, but also 
\begin{quote}
Queergaming engages different grammars of play, radical play, not grounded in normative ideologies like competition, exploitation, colonisation, speed, violence, rugged individualism, leveling up, and win states \cite{chang2017queergaming}
\end{quote}
Chang’s definition is not intended to be static or prescriptive, and for sure there are examples of queer play that lean into the conventions mentioned above, such as speedrunning \cite{rubergspeed}. There is a diverse range of rich scholarship on queering games in terms of alternate play styles and mechanics. There are two strands that are particularly relevant to the present discussion; power dynamics in queer games and archival play. Mattie Brice takes a different tack on the matter of empathy:
\begin{quote}
If we understand play as the exercising of empathy through engaging contexts, and kink as a type of play design that deeply confronts life contexts, then kink practises stand as a stronger model for engaging with meaningful play \cite{brice}
\end{quote}
Sihvonen and Harviainen \cite{sihvonenmy2023} also draw parallels between kink, BDSM, play and queer gaming communities, especially the performative and ritualistic aspects of each. Lander \cite{lander2019powergaming} has also done work on the political and embodied context of game-making that “exploit the boundary of the magic circle.” This work provides a different perspective on empathy, pleasure and games, and also centres conversations over personal and societal power dynamics.

Melissa Kagen’s term “archival adventuring,” refers to games which afford players to act in ways akin to an archival researcher \cite{kagenarchival2020}. This is relevant not only for the link with how some archaeogamers occupy and study video games, but also for the queer archival potential of games such as \textit{What Remains of Edith Finch} which constantly disorients the player by asking them to occupy the memories of different people across the game, along with remapped controls. Conversely, Pavlounis discusses how although \textit{Gone Home} also encourages an “archaeological understanding of history,” but only allows for a gated “straightening” of the game as archive, despite the queer subject matter \cite{pavlounis}.

Both archaeogaming and queer games studies have contended with the archival potential of games, just on different terms. Furthermore, both arguably involve engaging with forms of transgressive play that were not originally intended or foreseen by designers.

\section{Motivation}
The above background section is certainly not an exhaustive review of all literature relating to archaeogaming and queer game studies, and any such curation of sources will have gaps. Furthermore, to taxonomize these fields inevitably limits them, but it is hoped that by tracing potential parallels between them we can see how they are indirectly in conversation with each other. One motivation of this work was to demonstrate that both archaeogaming and queer games studies have been the result of community labour, in the case of the former this was a “grassroots initiative from people committed to two aspects of their lives in which they saw a valuable connection,” \cite{politopoulos2023finding} and for queer game studies that is “founded on a commitment to building bridges between theoretical analysis and the LGBTQ people who make, play and study games.” \cite{rubergfun} 

The remainder of this paper examines vignettes from the author’s own archaeogaming work, the motivation being to demonstrate the queer potential to be found through a close reading of each. Reflecting on this work allows for personal subjectivity and my lived experience as a queer researcher to be foregrounded. Stang \cite{stang} has discussed how close reading is a feminist and queer practice that is intimate, vulnerable, and pushes back against the reactionary critique that games should be ‘just for fun.’ 

\section{Vignettes}

\subsection{Wurm Online}
We conducted a “go-along” study in the MMO \textit{Wurm Online} \cite{smith2024}. The go-along is a combination of participant observation and interview, typically involving joining a participant on an everyday walk as a qualitative research method. Notably, Sam Stiegler's work on the embodied experience of go-alongs with trans, queer and non-binary youth \cite{stiegler2021} was a major influence on this work. While there are some limited examples of digital go-alongs \cite{go-along}, to date the method has not been explored in a video game context. We wanted to conduct a go-along in \textit{Wurm} because of the game’s particular affordances; \textit{Wurm Online} is a fantasy open-world sandbox MMO which allows players to have a permanent impact on the landscape, however structures will slowly decay over time unless they are regularly maintained or kept on a paid ‘deed.’ \textit{Wurm} was originally released in 2006 and its oldest intact server, Independence, dates to 2009. Some structures are at least a decade old and still persist in the game. One such infrastructure project is Dragon Fang Pass, a tunnel through the largest mountain on the Independence server that represents a huge amount of player time invested. We chose Dragon Fang Pass as the initial focus on our go-alongs as its is designated a heritage site by the \textit{Wurm} community, and thus would potentially be particularly well-suited to a methodology informed by personal memories of place.
The full results of this study, with the thematic analysis of the go along transcripts, are published separately\cite{smith2024}, for the purpose of this discussion a few relevant threads will be teased out. The very nature of the go-along as mediated around walking links with discussions around queer wanderings in games. Both Pelurson \cite{pelursonflanerie2019} and Kagen \cite{kagenarchival2020} have made the link between walking simulators and the figure of the flâneur, popularised in the 19th century as a cosmopolitan figure detached from society and able to reconfigure the space of the city according to their perambulations. Ruberg \cite{rubergspeed} also points out that:
\begin{quote}
The observation that gamers find walking simulators “boring” is linked to the basic premise on which calling a game a walking simulator is considered an insult: That the pace of walking is itself somehow bad or unacceptable.
\end{quote}
Conducting the go-alongs rendered \textit{Wurm} into a walking simulator, as we went on meandering paths dictated by our participants. Linking back with the topic of discomfort mentioned previously, there were multiple times where I found it difficult just to perform basic actions in the game, like climbing a ladder, due to my unfamiliarity with the controls on the Steam Deck. These moments were embarrassing but also led me to reflect on my role as a player-researcher and further underlined that the participants were the experts.
One participant, Nirav, is the curator of the in-game Rockcliff Museum. In \textit{Wurm}, certain items will carry the signature of their creator, but this can degrade over time and only be recovered through the improvement mechanic, which takes a considerable amount of time and resources:
\begin{quote}
Nirav: So many hours, and, and, and really most of the hours have been in improving items to get signatures, because to find the signature, you get another letter every ten [points]
\end{quote}
Improving signatures is a core part of Nirav’s curatorial practise, and it also an important part of community events, as was stressed by another participant:
\begin{quote}
P3: And impalongs, impalongs also will have some lag, because of all the players gathered together with uh, the game needing to keep track of light sources and animals and you know, things like that. Those give you more of an indica- more of a, I dunno, a feel, a sense for how it used to be in terms of when players used to be gathered together in a community.
\end{quote}
Imping has a kind of ritualistic aspect. In her book\textit{ Wandering Games} \cite{kagenwandering} Kagen reflects on the potential for rituals and wanderings to provide an outlet for those under societal constraints. This also chimes with an observation made by Nirav:
\begin{quote}
Nirav: I've put a lot of time into it, but it's been a lot of fun, and I mean, to be honest, I, I really believe that most of the hardcore players in this game are older and disabled…And I fall into that category, I'm 55. And I've had some medical life issues that have... enabled me to play this much, you know…And pour myself into something that, you know, is is pretty... pretty exciting and cool.
\end{quote}
Kara Stone \cite{stone} has written on her artistic practise as a queer disabled developer, specifically the potential for a reparative game design that:
\begin{quote}
...situated in the everyday. It is connected to repetition, mundanity, and urgency, rather than event, catastrophe, and emergency. I see it as a daily ritual, something that needs to be tended to everyday
\end{quote}
Imping items in \textit{Wurm} can be seen as a form of reparative curatorial design in the context of Rockcliff Museum,  and reparative play more broadly. The go-along method itself has queer potential in its focus on the apparently mundane act of walking and meandering through a familiar space, but also reconfiguring that space as a site of personal reflection with no specific goals or need for efficiency.

\subsection{Elden Ring}
In 2022, we conducted an archaeological survey of the action-adventure game \textit{Elden Ring} \cite{smith2022dark}. Like other games in the Soulsbourne genre, \textit{Elden Ring} has an asynchronous multiplayer which allows a player to interact with messages left in the landscape, as well as bloodstains denoting where another player died.  We focused on recording player messages and bloodstains at two locations, with the intention of testing if elements of the metagame and player experience could be inferred from them. More recent work directly contends with the queerness of \textit{Elden Ring} \cite{leiter}, in which Leiter suggests that it may be more productive to find evidence of the queer community through looking at fan archives and wikis then an archaeological survey of this kind. While we agree that those sources are invaluable, we would also like to demonstrate the queer potential of this methodology in \textit{Elden Ring}.

Soulsborne games are notorious for being difficult, and \textit{Elden Ring} is no different, however we encountered a different kind of friction when conducting the survey than simply dieing multiple times. We adapted existing archaeological recording techniques traditionally used in analogue space, such as photography and plan-making. In the case of the latter, in order to create scale base plans, I used my player character’s foot as a unit of measurement. This meant walking at tiny incremental steps around the periphery of the game locations that we surveyed. This process was incredibly laborious, but also allowed for a deeper understanding of how the space was designed. To reference the earlier discussion of power dynamics in games, on the one hand to archaeologically survey \textit{Elden Ring} is to engage with the game and its landscape in an unconventional way, on the other hand, we were still constrained by the algorithmic tendencies of the game’s server and what (and when) it would reveal more player-created content. Furthermore, because using mods carries a risk of being banned from the game’s multiplayer mode, we didn’t explore this as an option. That being said, as Chang puts it:
\begin{quote}
…imagine play, exploration, even failure that resists the check box, the Boolean, that rewrites protocol and refuses the posthuman fantasies of power and control \cite{chang2017game}
\end{quote}
In archaeology, the concept of ‘context’ is incredibly important; it relates to the position and associations of an artefact or feature, which helps determine its chronological relationship relative to other finds and features. For example, if a pottery sherd is found in dark soil that is distinct from a layer of lighter soil below it in which another sherd was deposited, the two sherds are found in different archaeological contexts and the darker soil is likely later. We attempted to record the spatial and temporal context of each message and bloodstain by drawing plans and making note of what day and at what observable number of server refreshes a particular asset first appeared. The concept of ‘context’ is also incredibly important in kink:
\begin{quote}
Video games are preoccupied with tech progressivism and late capitalistic practices that bank on ripping the sutures between reality and play. WE ARE ALWAYS PLAYING. WE ARE ALWAYS IN CONTEXTS. CONTEXT IS EVERYTHING. \cite{brice}
\end{quote}
The archaeological surveying of video games allows us to engage with their context and the power dynamics involved on a personal level. Arguably, this means the survey is actually a record of how we as researchers involved in the project experienced the game, as much as it is a record of player activity. As Graham puts it, “When I write about video games, I am the player-subject I know best.” \cite{graham2020} This could seem limiting, but from a game preservation point of you, it is invaluable to have a record of the experience of play. Pow has written about the potential for a reparative video game history:
\begin{quote}
The act of writing video game history gives us the possibility of reading history as something alive that churns, sets in motion, breathes life into, and animates the documents that have been saved (and those that have been lost), displacing the records that have come to stand in for the people they are attached to. \cite{pow2019outside}
\end{quote}
The queer potential of an archaeological survey of \textit{Elden Ring} is to engage with the personal context of its recording and the incidental traces that players leave behind in the game’s landscape.

\subsection{Nothing Beside Remains}

\textit{Nothing Beside Remains} (hereafter NBR) is a short exploration game which is set in the ruins of a village, with no set goals or win states. The village is procedurally generated through the process of an abstract simulation of a population that has the possibility to reach one of three failure conditions, such as ecosystem collapse, which effects the eventual algorithmic generation of structures and material culture. Informed by Cook’s work on generative forensic games \cite{cookinfo}, I have dubbed this a “generative archaeology game” \cite{smith2023darned} in that a player arguably is tasked with archaeologically interpreting the output of the generator. In a 2022 GDC talk \cite{gdc} I put forward the idea that environmental storytelling in games requires the player to archaeologically interpret a game environment, building on the work of Livingstone et al \cite{livingstone} who consider this to be a form of “archaeological storytelling.” Other archaeogaming scholars have also taken up the idea of the “archaeological mindset” \cite{caracciolo2022materiality} and “archaeological fandom” \cite{bennett2023lasts} in games. I believe this to be a burgeoning new strand for archaeogaming that moves beyond representation and provides the opportunity to engage more deeply with level and narrative design in games and their archaeological implications.

In 2023, we conducted a player study with NBR, providing a bespoke build of the game and an online questionnaire. We asked to participants to interpret what had happened to the village to see what reasons they gave for these interpretations on the basis of the implicit environmental storytelling in NBR. There were 187 participants in total, and the preliminary results of the study and their implications for archaeological storytelling have been published \cite{smith2023darned}. For the purpose of this discussion, I am interested in the queer affordances of NBR as a game with a procedurally generated setting.

In Anna Anthropy’s \textit{Queer’s In Love At The End Of The World}, a player has 10 seconds to interact with their lover before the world inevitably ends. Ruberg has argued that rather than this being a permadeath game, it can actually be classified as a \textit{permalife} game in which death is impossible because it can be endlessly looped. They explain:

\begin{quote}
At the most basic level, the mechanic of permalife could be seen as a powerful refusal of death – a symbolic performance of the will to live in the face of homophobic oppression and violence. Within the context of video games, permalife could also be read as a rebuttal, through design, to the masculinist ‘hardcore-ness’ of permadeath. \cite{rubergpermalife}
\end{quote}

NBR, as a game with procedurally generated content and an abstract ASCII aesthetic, consciously borrows from the roguelike genre, which is also characterised by the permadeath mechanic. However, it arguably is also a permalife game for several reasons. Firstly, there is actually no way to die in NBR, and secondly, because of its procedurally generated nature, it is always possible to regenerate a new village. The concept of permalife has interesting implications for how we might conceive of the archaeological record. Perry \cite{perryenchantment} has argued that the narrative of archaeology being a finite and non-renewable resources is damaging, cynical and inaccurate, urging us to instead cultivate enchantment with the archaeological record. She defines this as:
\begin{quote}
‘a state of wonder’ that typically entails surprise, pleasure, uncanniness (discomfiture), presence, or sensory agitation.
\end{quote}
If archaeology is seen as a destructive process and the archaeological record non-renewable, then this is akin to permadeath. An enchantment model of the archaeological record leaves room for possibilities beyond models of resource extraction and consumption, for permalife. This ties back in with NBR as a game that allows for endless archaeological speculation. It is important to state that enchantment, permalife and an endlessly regenerating ruined village do have uncomfortable implications. As Ruberg puts it: 
\begin{quote}
...what it means to go on living in these games is far messier and less utopian than it may at first appear. The futurity found in this work is, in turns, exuberant, weary, mournful and defiant. In contrast to the neo-liberal, homonormative narrative of LGBTQ lives and histories ‘getting better’ \cite{rubergpermalife}
\end{quote}
The permalife of the archaeological record is arguably represented in NBR, and though that provides endless opportunity for discovery, it also involves the endless looping of ruination and all that implies for this fictional village. Linking back with queer theory and Halberstam's work, we can imagine the infinite promise of failure in this cycle of permadeath-in-life. As Halberstam puts it:
\begin{quote}
Under certain circumstances failing, losing, forgetting, unmaking, undoing, unbecoming, not knowing may in fact offer more creative, more cooperative, more surprising ways of being in the world. Failing is something queers do and have always done exceptionally well. \cite{halberstam2011queer}
\end{quote}
The aesthetics of ruination in video games and the relationship between archaeogaming and queergaming are beyond the scope of this paper, but should be explored further in future work.

\section{Discussion and Conclusion}
Several themes have coalesced through reflecting on the queer potential of the three archaeogaming vignettes presented. Firstly, all three include some aspect of wandering as Kagen \cite{kagenwandering} would define it; the \textit{Wurm} go-along is explicitly characterised by walking and reflecting, the \textit{Elden Ring} work involved repeated wandering through the survey area and one of the main and only mechanics of NBR is exploration of the village. The gaming archaeologist perhaps, then, has some affinities with the flâneur and their reinscribing of space through how they inhabit it. The queer phenomenology of archaeogaming is definitely a fruitful avenue for further research.

Another theme is failure and frustration. In \textit{Wurm}, I had to contend with my own inability to successfully interact with objects like ladders, while \textit{Elden Ring} required repetitive and painstaking remediation of real-world surveying techniques into an environment not suited for them. With NBR, it is perhaps impossible to truly fail or die, but the uncanny existence of the procedural ruined village brings up disquieting implications.

In the case of \textit{Wurm}, reparative game design was relevant to the rituals that players conduct to both tend to the game world and themselves. A personal record can be an example of a reparative history with \textit{Elden Ring}, especially as it can account for the incidental traces that other players leave behind and may not think to take note of. Grinblat et al \cite{grinblatreparative} have theorised that narrative sandboxes like NBR allow for reparative play. By having deliberately incomplete narratives that invite a multiplicity of interpretations:
\begin{quote}
It’s more accurate to say these narratives arrive intact but buried... and that you’re given tools of revelation: a compass to find the relevant sites and a hammer and chisel to excavate them. 
\end{quote}
This archaeological metaphor is fitting given the reparative potential of generative archaeology games like NBR. Furthermore, the idea that these games “invite repair by arriving in disrepair” \cite{grinblatreparative} rhymes with Kagen’s sentiment that:
\begin{quote}
To clean up the gap-filled messiness of the archive by giving it one, true narrative – as game designers often do in the interest of telling a compelling story – is to unqueer it. \cite{kagenarchival2020}
\end{quote}
Which should also be read alongside Perry’s thoughts on enchantment:
\begin{quote}
In fact, archives themselves are premised on the notion that we can forever discover new and different things about their contents—hence the preservation of these contents, ripened for reinterpretation over time.
\end{quote}
The permalife of the archive is in its ability to continually be remade, reinterpreted, remediated. In the case of archaeogaming, embracing the personal, uncomfortable and reparative potential of the records we make allows us to see that, indeed, archaeogaming has always been queer.

\section{Postscript}

This paper is itself an attempt at an archive of archaeogaming and queergaming literature, and for that reason, it is a failure. No archive is ever complete. However, through exploring these vignettes on my own personal experience of archaeogaming research, I hope to have shown the queer potential of adapting archaeological methodologies and concepts in the context of video games. I follow Pow’s work on \textit{A Trans Historiography of Glitches and Errors}:
\begin{quote}
Re-centering the ephemeral, and the acts, erasure, and embodiment within computer and video game history, allows us to argue that what has been lost—the objects missing from archives, the trans histories that are left out of historical narra-tives, all things that have not been recorded—have a life and a history that we can still trace, and which is still meaningful\cite{pow2021trans} 
\end{quote}
This sentiment is in conversation with Aycock and Biittner’s work on experimental archaeogaming, and their concerns about:
\begin{quote}
...the invisibility of the production process in the (digital) archaeological record... although we may have the end product of the video game production sequence (i.e., the game), the tools used can be invisible or “lost” in the assemblages we are working with\cite{aycock2024experimental}
\end{quote}
While archaeogaming can learn from the centring of the personal, embodied experience, queergaming could benefit from experimental archaeogaming techniques in the pursuit of tracing queer and trans media histories.

\begin{acks}
Thanks to Mike Cook, who collaborated with me on the three projects referenced as vignettes in this paper. 

I would like to thank organisers and attendees of the 2023 Hunt-Simes Institute in Sexuality Studies at the Sydney Social Sciences and Humanities Research Centre who encouraged me to continue to pursue queer studies of archaeogaming.

My sincere thanks to the reviewers for their insightful suggestions and feedback which I have done my best to include here. 

I would also like to thank my past self for persevering with independent research and presenting the original paper \textit{Archaeogaming as Queegaming} at the CAA conference in 2018 at a time when it felt very vulnerable for me to do that work.

This work was supported by the EPSRC Centre for Doctoral Training in Intelligent Games  \&  Games Intelligence (IGGI) [EP/L015\-846/1].
\end{acks}

\bibliographystyle{ACM-Reference-Format}
\bibliography{biblio,Archaeogaming,Queergaming,software}

\end{document}